\documentclass[aps,reprint,aps,showpacs,twocolumn]{revtex4-1}
\usepackage[usenames]{color}
\usepackage{bm}
\usepackage{amsmath}
\usepackage{amssymb}
\usepackage{epsfig}

\newcommand{\bee}{\begin{eqnarray}}
\newcommand{\ene}{\end{eqnarray}}

\newcommand{\ran}{\rangle}
\newcommand{\eqb}{\begin{equation}}
\newcommand{\eqe}{\end{equation}}

\newcommand{\vare}{\varepsilon }
\newcommand{\pd}{\partial}

\begin{document}

\title{The three-site Bose-Hubbard model subject to  atom losses:    the boson-pair
dissipation channel and failure of the mean-field approach}

\author{ V. S. Shchesnovich$^{1}$ and D. S. Mogilevtsev$^{2,3}$  }

\affiliation{${}^1$Centro de Ci\^encias Naturais e Humanas, Universidade Federal do
ABC, Santo Andr\'e,  SP, 09210-170 Brazil\\ ${}^2$Institute of Physics, Belarus
National Academy of
Sciences, F.Skarina Ave. 68, Minsk 220072 Belarus; \\
${}^3$Instituto de F\'{\i}sica, UNICAMP, CP 6165, Campinas-SP, 13083-970, Brazil}

\begin{abstract}

We employ the  perturbation series expansion  for derivation of the reduced master
equations for the three-site Bose-Hubbard model subject to strong atom losses from
the central site. The model  describes a condensate trapped in a triple-well
potential subject to externally controlled removal of atoms.  We find that the
$\pi$-phase state of the coherent superposition between the side wells decays via
two dissipation channels, the single-boson channel (similar to the externally
applied dissipation) and the boson-pair channel. The quantum derivation is compared
to the classical adiabatic elimination within the mean-field approximation. We find
that the boson-pair dissipation channel is not captured by the mean-field model,
whereas the single-boson   channel is described by it. Moreover, there is a
matching condition between the zero-point energy bias of the side wells and the
nonlinear interaction parameter which separates the regions where either the
single-boson or the boson-pair dissipation channel dominate.  Our results indicate
that the $M$-site Bose-Hubbard models, for $M>2$, subject to atom losses may
require an analysis which goes beyond the usual  mean-field approximation for
correct description of their dissipative features. This is an important result in
view of the recent experimental works on the single site addressability of
condensates trapped in optical lattices.

\end{abstract}
 \maketitle

\section{Introduction}

The mean-field approximation, formally obtained by  replacing the boson creation
and annihilation  operators by complex scalars, is usually employed for description
of bosonic many-body systems when the number of bosons is large (for instance, see
Refs. \cite{BStat,CD,GP}). Such a replacement can be  justified by the
$\sqrt{N}$-scaling of the boson operators, that is, when the populations are large,
the commutators --  the source of quantum corrections -- are negligible
\cite{BStat}. The relation between the  quantum and mean-field descriptions is a
subject of intensive studies. The quantum description is necessary for the
bifurcations, which modify significantly the quantum spectrum \cite{AFKO},  the
quantum collapses and revivals \cite{MCWW}, and    the many-body quantum
corrections to the mean-field theory \cite{VA,AV}. Making explicit the $N$-scaling
of the operators and identifying the $N$-scaling of the parameters for a fixed
particle density, reveals the link of the mean-field approximation to the
Wentzel-Kramers-Brillouin  semiclassical approach to the discrete Schr\"odinger
equation \cite{Braun}, now employed in the Fock space with the inverse number of
bosons $1/N$ playing the role of an effective Planck constant (see, for example,
Ref. \cite{ST}). Therefore, the mean-field limit, as the semiclassical limit of a
discrete Schr\"odinger equation, is also \textit{singular}. Hence, besides the
pronounced quantum corrections/fluctuations at the bifurcations/instabilities, one
must be prepared to find even a \textit{qualitative} disagreement between the
mean-field description and the full quantum consideration even when the populations
are large, as it is the case, for instance,  in the nonlinear boson model which
describes tunneling of boson pairs between two modes, see  Refs. \cite{SK,MsQ}.

\begin{figure}[htb]
\begin{center}
\epsfig{file=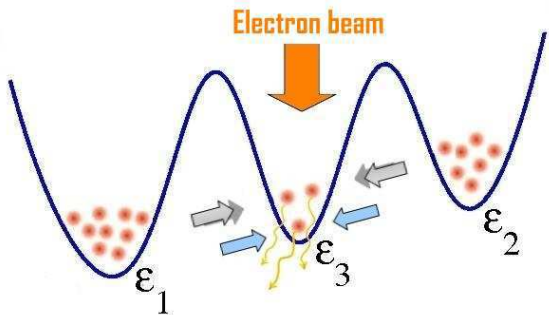,width=0.45\textwidth} \caption{(Color online) The three-site
Bose-Hubbard model setup. A laser/electron beam  removes atoms from the central
well with a rate $\Gamma$. The $\vare_j$ represents the zero-point energy of the
$j$th well. The two quantum channels of dissipation of the $\pi$-phase coherent
mode, described in the text, are shown schematically by arrows.  }
\label{FG1}
\end{center}
\end{figure}

The main purpose of the present paper is to study the dissipation dynamics of the
atom filled lattice sites coupled to a common dissipated sites. Our motivation is
the recent advancement in the quantum microscopy techniques and the current
experiments on the single site addressability in the optical lattices
\cite{SiteAddr1,SiteAddr2}, where controlled atom losses are induced in selected
sites of a two-dimensional optical lattice. We develop a direct method based on the
perturbation expansion for derivation of the reduced quantum master equations for
the Bose-Hubbard models with dissipation (we consider the atom losses) and  compare
the quantum derivation with the mean-field description. We concentrate on the
three-site Bose-Hubbard model, which  is the  simplest model describing atom filled
sites coupled to a common dissipated one and describes, for instance,  cold atoms
or Bose-Einstein condensate trapped in a triple-well potential subject to removal
of atoms from the central well, see Fig. \ref{FG1}. The three-site Bose-Hubbard
model was also noted to possess many features of complexity of a general quantum
dynamics, as the Wigner-Dyson spectral statistics and quantum chaos
\cite{WDSt,Chaos3w}. It is also the simplest model where the boson-pair tunneling,
originating from the nonlinearity of the model,  is possible.

We derive  the reduced quantum  master equations  for the coherent modes describing
the condensate in the side wells, then consider   the  mean-field approach and
compare the results of the two approaches. We note here that we consider an open
quantum system and, as such, described by the quantum master equation
\cite{CBook,BP}. However, in the case of the  atom losses, the mean-field
formulation is straightforward (this also applies to the case of the multiple-site
atom losses of the experimental works of Refs. \cite{SiteAddr1,SiteAddr2}).
Contrary to the fact that the mean-field approximation applies with a good accuracy
to the \textit{two-site} Bose-Hubbard model with   atoms losses and a noise
\cite{TWW,NlZeno},  it is shown here  that the correct and complete description of
the three-site model (in general, the $M$-site Bose-Hubbard models, with $M\ge3$)
requires going beyond the usual  mean-field approach. This disagreement stems from
the quantum boson-pair dissipation channel, due to the nonlinear interaction (this
is similar to the boson-pair tunneling resulting in the qualitative failure of the
mean-field approach in Ref. \cite{MsQ}). Moreover, there is a matching condition
between the zero-point energy bias of the side wells and the nonlinear interaction
parameter which separates the regions where either the single-boson or the
boson-pair dissipation channel dominate. Hence, one has to use the full quantum
consideration, i.e. the quantum master equation reduction methods, to describe the
decay of  the subsystem (in our case, the quantum modes describing the side wells),
which then may be treated with further approximations, even \textit{resembling} the
mean-field approach. However, the point is that without invoking the  quantum
consideration at some  stage, i.e. working just within  the mean-field approach,
one will be unable to describe the  dissipative behavior of the filled sites
coupled to a common dissipated site, which conclusion  is  also relevant  to the
recent experiments of Ref. \cite{SiteAddr1}.

The paper is organized as follows. In section \ref{sec1} we formulate the quantum
master equation. The derivation of the reduced master equation for the side modes
(i.e., the coherent superposition modes over the left and right wells in Fig.
\ref{FG1}) is given in section \ref{sec2}. In section \ref{sec3} a similar
reduction is applied to the reduced master equation of  section  \ref{sec2}
producing the master equation for the $\pi$-phase coherent mode. In section
\ref{sec4} the adiabatic elimination within the mean-field approximation is
studied. The concluding remarks and discussion is contained in section \ref{sec5}.

\section{The problem formulation in terms of the  master equation }
\label{sec1}

A quantum channel representation  for a local  atom losses (e.g., from a single
lattice site), see Fig. \ref{FG1}, can be given in the Fock space as follows
\cite{NlZeno}
\[
|k_j\rangle|0\ran_R\to\sqrt{p}|k_j-1\rangle|1\ran_R+\sqrt{1-p }|k_j\rangle|0\ran_R,
\]
$|k_j\rangle$ is the ket-vector of the Fock space of a dissipating  boson mode,
$|X\ran_R$ describes the atom counter, and $p=p(k_j,t)$ is the probability. Note
that $p(k_j,\delta t)$, for small $\delta t$, depends linearly on the number of
atoms in the dissipating mode. As the result,  a Lindblad term  with the generator
$L = \sqrt{\Gamma}a_j$ appears in the master equation for the density matrix. Here
$\Gamma$ is the dissipation rate parameter. We consider the dissipation acting on
the central well (denoted with $j=3$ in Fig. \ref{FG1}) of the triple-well trap,
thus the master equation for the density matrix reads
\eqb
\frac{d \rho}{dt} = -\frac{i}{\hbar}[H,\rho] + \Gamma\left(a_3\rho a^\dag_3 -
\frac12n_3\rho -\frac12\rho n_3\right).
\label{EQ1}\eqe
By $H$ we denote  the three-site Bose-Hubbard Hamiltonian,
\eqb
H = -J ([a_1^\dag +a_2^\dag]a_3 + a_3^\dag [a_1+a_2]) +
\sum_{j=1}^3\vare_ja^\dag_ja_j + U \sum_{j=1}^3 (a^\dag_j)^2a_j^2,
\label{EQ2}
\eqe
where  $a_j$ and $a_j^\dag$ ($j=1,2,3$) are the local boson modes describing
occupation of the respective  well, $J$ is the linear tunneling rate, $\vare_j$ is
the zero-point energy of the respective well, and $U$ is the atomic interaction
parameter proportional to the $s$-wave scattering length.

Due to  the linear coupling of the  central well to the side wells in the
Bose-Hubbard model, it is convenient to use the new canonical basis $b_1 =
(a_1+a_2)/{\sqrt{2}}$, $b_2 =(a_1-a_2)/{\sqrt{2}}$, and $b_3 = a_3$. Here  the
modes $b_{1,2}$ describe, respectively,  the zero-phase and $\pi$-phase coherent
superpositions between the side wells.  The Bose-Hubbard Hamiltonian becomes
\begin{eqnarray}
H &=& -J_{13}(b^\dag_1 b_3+b^\dag_3b_1) - J_{12}(b^\dag_1b_2+b^\dag_2b_1) + \vare
n_3 \nonumber\\
&+ & Un_3^2 + \frac{U}{2}\left( [n_1+n_2]^2 + [b^\dag_1b_2+b^\dag_2b_1]^2\right),
\label{EQ3}\end{eqnarray}
where $n_j = b^\dag_jb_j$ and we have introduced the parameters $J_{13} =
\sqrt{2}J$, $J_{12} = (\vare_2-\vare_1)/2$, $\vare = \vare_3 -(\vare_1+\vare_2)/2$
(and dropped an inessential term proportional to $N=n_1+n_2+n_3$).

\section{The reduced master equation for the   side wells}
\label{sec2}
In the strongly dissipative case, $\Gamma\gg J_{13}/\hbar$,  the central well is
emptied on the time scale $t\sim 1/\Gamma$, while the coherent modes $b_1$ and
$b_2$ almost retain their populations (see also Ref. \cite{NlZeno}). It turns out
that in this case one can derive an approximate master equation for the side wells
alone valid for the times $t\gg 1/\Gamma$, when the central-well dissipation  has
the fastest time-scale in the system. Introducing the small parameter $\epsilon =
J_{13}/\hbar\Gamma$ we require that $J_{12}/\hbar\Gamma = \mathcal{O}(\epsilon)$
and $U\langle n_{1,2}\rangle/\hbar\Gamma = \mathcal{O}(\epsilon)$. This
approximation is valid for an arbitrary bias $\vare$ between the central and side
wells.  We will see that, already in the first order in $\epsilon$, the density
matrix for the whole system is not factorized and the central-well population
strongly depends on that of the side wells. Nevertheless, in the higher orders of
the perturbation theory (namely, in the second and third orders in $\epsilon$), the
central well acts as an effective reservoir for the side wells leading to the
master equation in the standard Lindblad form.

First of all, let us pass to the interaction picture. Introduce
\begin{eqnarray}
&& \hat{H}_{I}(t) = U_0^\dag(t) H_I U_0(t),\quad \hat{\rho}= U_0^\dag (t)\rho(t)
U_0(t)\nonumber\\
&& U_0(t) = U_{12}(t)\otimes U_{3}(t)=\exp\{-\frac{i}{\hbar}(H_{12}\! +\!H_3)\Delta
t\},
\label{EQ19}\end{eqnarray}
where $\Delta t = t-t_0$, $t_0$ is some initial time, and $H_{12}$, $H_{3}$ and
$H_I$ are parts of the Hamiltonian (\ref{EQ3}) describing the side wells, the
central well and the interaction between them, respectively (we have also
subtracted the nonessential term $U(n_1+n_2+n_3)$ from the system Hamiltonian to
simplify the presentation). The master equation in the interaction picture reads
\eqb
\frac{d\hat{\rho}}{dt} = -\frac{i}{\hbar}[\hat{H}_I(t),\hat{\rho}]+\Gamma
\left(\hat{b}_3(t)\hat{\rho} \hat{b}^\dag_3(t) - \frac12\hat{n}_3\hat{\rho}
-\frac12\hat{\rho} \hat{n}_3\right),
\label{EQ20}\eqe
where  $\hat{b}_3(t) = U_0^\dag(t)b_3U_0(t)$. \textit{Below we will work in the
interaction picture and  drop the hat, for simplicity.} The density matrix
expansion reads $\rho = \rho^{(0)}_{12}\otimes\rho^{(0)}_3 + \rho^{(1)} +
\rho^{(2)} + \mathcal{O}(\epsilon^3)$, where, for instance, $\rho_{12} =
\mathrm{Tr}_{3}\{\rho\}$ (an arbitrary non-factorized expansion, $\rho^{(0)} =
\sum_{ij} c_{ij}\rho^{(0)}_{12i}\otimes\rho^{(0)}_{3j}$, leads to the same result).
In the lowest orders in $\epsilon$ we have:
\eqb
\frac{d \rho^{(0)}_{12}}{dt} =0,\quad  \frac{d \rho^{(0)}_3}{dt} = \Gamma
D(t)\{\rho_3^{(0)}\},
\label{EQ21}\eqe
\eqb
\frac{d \rho^{(1)}}{dt} =
-\frac{i}{\hbar}[H_I(t),\rho^{(0)}_{12}\otimes\rho^{(0)}_3]+\Gamma
D(t)\{\rho^{(1)}\},
\label{EQ22}\eqe
\eqb
\frac{d \rho^{(2)}_{12}}{dt} =
-\frac{i}{\hbar}\mathrm{Tr}_{3}\{[H_I(t),\rho^{(1)}]\}.
\label{EQ23}\eqe
The solution to the  equation for $\rho^{(0)}_3$ is as follows
\eqb
\rho^{(0)}_3(t) = |0\rangle\langle0|+e^{-\frac{\Gamma}{2}\Delta
t}\left[c_1|1\rangle\langle0|+ h.c.\right]+ \mathcal{O}(e^{-\Gamma\Delta t}),
\label{EQ24}\eqe
where the last term represents all higher-order external products of the Fock basis
states and a decaying renormalization correction to the first term. The coefficient
$c_1$ has the following meaning: $\mathrm{Tr}_3\{ b_3(t)\rho_3^{(0)}(t)\} =
c_1f_1(t)e^{-\frac{\Gamma}{2}\Delta t}+ \mathcal{O}(e^{-\Gamma\Delta t})$, with
$f_1$ being  the eigenvalue of the unitary transformation in Eq. (\ref{EQ20}):
$U_3(t)|1\rangle = f_1(t)|1\rangle$, in our case $f_1(t) = e^{-i\vare\Delta
t/\hbar}$ (note also that $b_3(t) = f_1(t)b_3$).

Inserting Eq. (\ref{EQ24}) into Eq. (\ref{EQ22}) we get the general solution in the
following form  (suggested by the form of $\rho^{(0)}_3$ itself)
\eqb
\rho^{(1)}(t) = V(t)\left[\rho^{(0+1)}_{12}\otimes|0\rangle\langle0|
\right]V^\dag(t) +  \mathcal{O}(e^{-\frac{\Gamma}{2}\Delta t}),
\label{EQ25}\eqe
where $\rho^{(0+1)}_{12}=\rho^{(0)}_{12}+\rho^{(1)}_{12}$ and $V(t)$ reads
\eqb
V(t) = \exp\{\alpha_1(t)b_1(t)b^\dag_3(t)\}=I + \alpha_1(t)b_1(t)b^\dag_3(t)
+\mathcal{O}(\epsilon^2),
\label{EQ26}\eqe
with a  scalar function $\alpha_1(t)=\mathcal{O}(\epsilon)$. Indeed, taking
derivative of the solution (\ref{EQ26}) and using Eqs. (\ref{EQ22}) and
(\ref{EQ24}), we obtain the equation to be satisfied
\[
\frac{d \rho^{(1)}}{dt} = \frac{d V}{dt} \rho^{(0)}_{12}\otimes|0\rangle\langle0|
V^\dag + V \rho^{(0)}_{12}\otimes|0\rangle\langle0| \frac{d V^\dag}{dt}
\]
\[
+ V \frac{d \rho^{(1)}_{12}}{dt}\otimes|0\rangle\langle0| V^\dag=
-\frac{i}{\hbar}[H_I(t),\rho^{(0)}_{12}\otimes|0\rangle\langle0|]
\]
\eqb
+\Gamma
D\{V\rho^{(0+1)}_{12}\otimes|0\rangle\langle0|V^\dag\}+\mathcal{O}(\epsilon^2).
\label{EQ27}\eqe
We evaluate
\[
-\frac{i}{\hbar}[H_I(t),\rho^{(0)}_{12}\otimes|0\rangle\langle0|]\qquad \qquad
\]
\eqb
 =\frac{iJ_{13}}{\hbar}\bigl(f^*_1(t)b_1(t)\rho^{(0)}_{12}\otimes|1\rangle\langle0|-h.c.\bigr)
\label{EQ28}\eqe
and, using Eq. (\ref{EQ26}),
\begin{eqnarray}
&& \Gamma D(t)\{V\rho^{(0+1)}_{12}\otimes|0\rangle\langle0|V^\dag\}\qquad\qquad
\nonumber\\
&& = -\frac{\Gamma}{2}\left(\alpha_1(t)f^*_1(t)b_1(t)\rho^{(0)}_{12}\otimes|1\rangle\langle0|+
h.c.\right) +\mathcal{O}(\epsilon^2).\qquad
\label{EQ29}\end{eqnarray}
Eqs. (\ref{EQ22})  and (\ref{EQ28}) give immediately
\eqb
\frac{d{\rho}^{(1)}_{12}}{dt}=\mathcal{O}(e^{-\frac{\Gamma}{2}\Delta t}).
\label{EQ30}\eqe
Since  $J_{12}/\hbar\Gamma = \mathcal{O}(\epsilon)$ and $U\langle
n_{1,2}\rangle/\hbar\Gamma = \mathcal{O}(\epsilon)$, the derivative of $b_1(t)$ is
of order $\epsilon$,
\eqb
\frac{d b_1(t)}{dt}=-\frac{i}{\hbar}[H_{12},b_1(t)] = \mathcal{O}(\epsilon).
\label{EQ32}\eqe
Now, by taking into account  Eqs. (\ref{EQ26}), (\ref{EQ28})-(\ref{EQ32}),  one
sees that Eq.~(\ref{EQ27}) is satisfied  by setting
\eqb
\frac{d\alpha_1}{dt} = -\left[\frac{\Gamma}{2}+i\frac{\vare}{\hbar}\right]\alpha_1
+ \frac{iJ_{13}}{\hbar}.
\label{EQ31}\eqe
Equation (\ref{EQ31}) gives
\eqb
\alpha_1 =
\frac{2iJ_{13}}{\hbar\Gamma}\left[1+\frac{2i\vare}{\hbar\Gamma}\right]^{-1} +
\mathcal{O}(e^{-\frac{\Gamma}{2}\Delta t}).
\label{EQ33}\eqe
The final step is to insert Eq. (\ref{EQ25}) into Eq. (\ref{EQ23}), use Eqs.
(\ref{EQ21}) and (\ref{EQ30}) and take the trace over the central-well subspace,
keeping only the  terms up to $\mathcal{O}(\epsilon^2)$. Returning to the
Schr\"odinger picture, we get (up to a correction of the order
$\mathcal{O}(\epsilon^3)$)
\begin{eqnarray}
\frac{d{\rho}_{12}}{dt}&=& -\frac{i}{\hbar}[H_{12},
{\rho}_{12}]-\frac{i}{\hbar}[\vare_R b^\dag_1
b_1, {\rho}^{(0)}_{12}] +\Gamma_R D_1\{  {\rho}^{(0)}_{12}\}\nonumber\\
&&
+\mathcal{O}(e^{-\frac{\Gamma}{2}\Delta t}),
\label{EQ34}\end{eqnarray}
with $\Gamma_R$ given by
\eqb
\Gamma_R =|\alpha_1|^2\Gamma =
\frac{4J_{13}^2}{\hbar^2\Gamma}\left[1+\left(\frac{2\vare}{\hbar\Gamma}\right)^2\right]^{-1},
\label{EQ18}\eqe
and \mbox{$D_1\{ {\rho}\} = b_1 {\rho} b^\dag_1 - \frac{1}{2}\{b^\dag_1 b_1,
{\rho}\}$}. The interaction-induced  Lamb shift $\vare_R$ of the zero-point energy
of the coherent zero-phase mode $b_1$ reads
\eqb
\vare_R=-J_{13}\mathrm{Re}(\alpha_1) = -\vare\frac{\Gamma_R}{\Gamma}.
\label{EQ35}\eqe


Our aim is to further reduce Eq. (\ref{EQ34}) to a master equation  for the
coherent $\pi$-phase mode $b_2$, which has  unusual dissipation features (see
below). To this end, however, one has to consider the contribution to Eq.
(\ref{EQ34}) coming from the next order in $\epsilon$, i.e.
$\mathcal{O}(\epsilon^3)$. Indeed, similar to the derivation of this section, the
reduced master equation for mode $b_2$ is obtained under the conditions that
$J_{12},U\langle n_{1,2}\rangle\ll\hbar\Gamma_R$, what makes the Hamiltonian part
in the master equation (\ref{EQ34}) smaller than   the Lindblad part, hence the
former should be discarded in the present order $\mathcal{O}(\epsilon^2)$. Thus, in
the second-order approximation, the coherent mode $b_2$ has no dissipation dynamics
at all (only the Hamiltonian evolution described by $H_2 = \frac{U}{2}n_2^2$).  It
only  appears in the higher-order version of the  master equation (\ref{EQ34}).

To derive the third-order correction to Eq. (\ref{EQ34}), we need to find the form
of $\rho^{(2)}(t)$, which satisfies equation similar to Eq. (\ref{EQ22}), but now
with the inhomogeneous term
\begin{eqnarray}
&&-\frac{i}{\hbar}[H_I(t),V\rho^{(0+1)}_{12}\otimes|0\rangle\langle0|V^\dag]
\nonumber\\
&&=\frac{iJ_{13}}{\hbar}\biggl\{\left(\sqrt{2}\alpha_1(t) f_2^*(t)b_1^2(t)\rho_{12}^{(0)}\otimes|2\rangle\langle0|-h.c.\right)
\nonumber\\
&&
-2i\mathrm{Im}\{\alpha_1(t)\}b_1(t)\rho_{12}^{(0)}b_1^\dag(t)\otimes|1\rangle\langle1|
\nonumber\\
&&+\left(\alpha_1(t) b_1^\dag(t)b_1(t)\rho^{(0)}_{12}-h.c.\right)\otimes
|0\rangle\langle0| \nonumber\\
&&+\left(f_1^*(t)b_1(t)\rho_{12}^{(1)}\otimes|1\rangle\langle0|-h.c.\right)
\biggr\}+\mathcal{O}(\epsilon^2),
\label{EQ36}\end{eqnarray}
where $U_3(t)|2\rangle = f_2(t)|2\rangle$, i.e. $f_2 = e^{-2i\frac{(\vare
+U)}{\hbar}\Delta t}$. The first three lines of Eq. (\ref{EQ36}) give the terms
additional to those in Eq. (\ref{EQ28}). Expression (\ref{EQ36}) also defines the
general form of $\rho^{(2)}(t)$:
\[
\rho^{(2)} = B_{00}(t)\otimes|0\rangle\langle0| +
B_{11}(t)\otimes|1\rangle\langle1|
\]
\eqb
 +\left(B_{10}(t)\otimes|1\rangle\langle0|+B_{20}(t)\otimes|2\rangle\langle0|+h.c.\right),
\label{EQ37}\eqe
where the operators $B_{ij}(t)=\mathcal{O}(\epsilon^{2})$ act on the subspace of
the side wells. They satisfy, in view of Eqs. (\ref{EQ18}), (\ref{EQ28}),
(\ref{EQ30}), (\ref{EQ35}) and (\ref{EQ36}), the following equations:
\begin{subequations}
\label{EQ39}
\begin{eqnarray}
\frac{dB_{11}}{dt} &=& -\Gamma B_{11}+ \Gamma_Rb_1(t)\rho_{12}^{(0)}b_1^\dag(t),
\label{EQ39a}\\
\frac{dB_{00}}{dt} &=& \Gamma B_{11} -\frac{i}{\hbar}[\vare_R
n_1(t),\rho_{12}^{(0)}]
\nonumber\\
&&-\frac{\Gamma_R}{2}\left(n_1(t)\rho_{12}^{(0)}+\rho_{12}^{(0)}n_1(t)\right),
\label{EQ39b}\\
B_{10} &=& \alpha_1(t)f_1^*(t)b_1(t)\rho_{12}^{(1)},
\label{EQ39c}\\
 B_{20} &=& \alpha_2(t)f^*_2(t)b_1^2(t)\rho_{12}^{(0)},
\label{EQ39d}
\end{eqnarray}
\end{subequations}
where $\alpha_1(t)$ is given by Eq. (\ref{EQ33}) and
\eqb
\frac{d\alpha_2}{dt} =
-\left[\frac{\Gamma}{2}+i\frac{\vare+U}{\hbar}\right]\alpha_2 +
i\frac{\sqrt{2}J_{13}}{\hbar}\alpha_1(t).
\label{EQ40}\eqe
One can easily solve  Eq. (\ref{EQ39a}) and by using integration by parts represent
the result as
\eqb
B_{11}(t) =
\frac{\Gamma_R}{\Gamma}b_1(t)\rho^{(0)}_{12}b_1^\dag(t)+\mathcal{O}(\epsilon^3)+\mathcal{O}(e^{-\Gamma\Delta
t}).
\label{EQ41}\eqe
Substituting Eq. (\ref{EQ41}) into Eq. (\ref{EQ39b}) and using the master equation
(\ref{EQ34}) one obtains
\eqb
B_{00}(t) = \rho_{12}^{(2)} +\mathcal{O}(e^{-\Gamma\Delta t}).
\label{EQ42}\eqe
Finally, the solution to Eq. (\ref{EQ40}) reads
\begin{eqnarray}
\alpha_2(t) &=&
-\frac{2\sqrt{2}J^2_{13}}{\hbar^2\Gamma^2}\left[1+i\frac{2\vare}{\hbar\Gamma}\right]^{-1}
\left[1+i\frac{2(\vare+U)}{\hbar\Gamma}\right]^{-1}
\nonumber\\
&+&\mathcal{O}(e^{-\frac{\Gamma}{2}\Delta t}) =
\frac{\alpha_1^2(t)}{\sqrt{2}}+\mathcal{O}(\epsilon^3)+\mathcal{O}(e^{-\frac{\Gamma}{2}\Delta
t}).
\label{EQ43}\end{eqnarray}

The fact that the operators $B_{11}$ and $B_{20}$ and $B_{02}$ do not contribute to
the equation for $\rho_{12}$ and the explicit expressions for $B_{00}$ and
$B_{10}$, Eqs. (\ref{EQ41}) and (\ref{EQ42}), lead to \textit{the same} Lindblad
form of the reduced master equation, i.e. in the Schr\"odinger picture we get
\eqb \frac{d\rho_{12}}{dt}=
-\frac{i}{\hbar}[H_{12}+ \vare_R n_1,\rho_{12}] + \Gamma_RD_1\{ \rho_{12}\}
+\mathcal{O}(e^{-\frac{\Gamma}{2}\Delta t}),
\label{EQ45}\eqe  
which is now valid up to correction of order $\mathcal{O}(\epsilon^4)$. Eqs.
(\ref{EQ37}), (\ref{EQ41})-(\ref{EQ43}) also show that the density matrix of the
full system $\rho$  can be rewritten as (now in the Schr\"odinger picture)
\eqb
\rho(t) = V\left[\rho_{12}(t)\otimes|0\rangle\langle0| \right]V^\dag +
\mathcal{O}(\epsilon^3)+ \mathcal{O}(e^{-\frac{\Gamma}{2}\Delta t}),
\label{EQ44}\eqe
where  $\rho_{12}(t)$ is taken up to the second order in $\epsilon$, $V(t) =
\exp\{\alpha_1 b_1b^\dag_3\}=I +\alpha_1 b_1b^\dag_3 + \frac12
\alpha_1^2b^2_1(b^\dag_3)^2$ with $\alpha_1$ given by Eq. (\ref{EQ33}). Eqs.
(\ref{EQ45}) and (\ref{EQ44}) are the main results of this section. Obviously, the
full density matrix (\ref{EQ44}) is not factorized, nevertheless, the reduced
density matrix of a subsystem satisfies the Markovian master equation (\ref{EQ45})
in the Lindblad form. Note that the difference between the density matrix
$\rho_{12}$ and the one obtained by tracing the full density matrix of Eq.
(\ref{EQ44}) is a constant term of the second order given by $B_{11}$ in Eq.
(\ref{EQ41}), which does not contribute to Eq. (\ref{EQ45}). Hence, Eq.
(\ref{EQ44}) is consistent with the approximation made.

\begin{figure}[htb]
\begin{center}
\epsfig{file=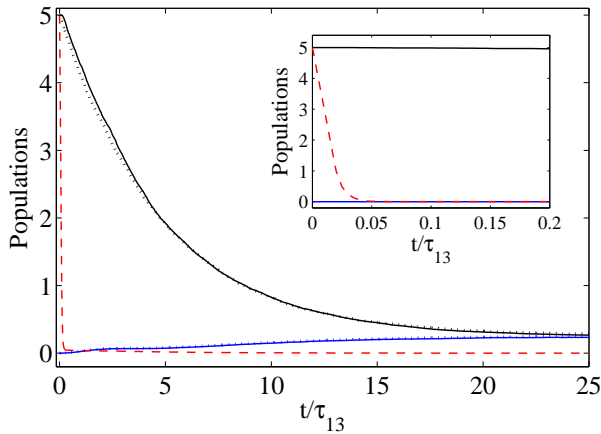,width=0.45\textwidth} \caption{(Color online) Comparison
between  the full 3-mode  Eq. (\ref{EQ1}) and the reduced 2-mode Eq. (\ref{EQ45}).
The population of $a_3$ is given by the dashed lines. The populations of modes
$a_1$ and $a_2$ (solid lines) of the 3-mode model are compared to that of the
2-mode model (dotted lines). Here $\Gamma=5J/\hbar$, $U=0.2J$, $\vare_j=0$. We have
used 5000 quantum trajectories. The inset  shows the short-time dynamics of
populations for larger $\Gamma=50J/\hbar$. The initial state is the Fock state with
the total of 10 atoms (5 occupying the mode $a_1$ and 5 occupying the mode $a_3$.}
\label{FG2}
\end{center}
\end{figure}

In Fig. \ref{FG2}, we use the Monte Carlo wave-function method \cite{MCD} and find
that   an excellent agreement of the reduced master equation (\ref{EQ45}) with the
full Eq. (\ref{EQ1}) extends also to the intermediate values of $\Gamma$ (we have
there $\epsilon = 0.2$). We have also verified that, for large $\Gamma$, the modes
$a_{1,2}$ (and, hence, $b_{1,2}$) stay practically unchanged while the dissipating
mode $b_3=a_3$ is quickly emptied  (the inset of Fig. \ref{FG2}).

Let us note the specific features of Eq. (\ref{EQ45}). We see that in the strongly
dissipative case, quite similar to Eq. (\ref{EQ1}), mode $b_2$ can retain a
significant part of its population, while $b_1$ loses almost all atoms, on the time
scale $t\sim 1/\Gamma_R$, and after that stays practically empty.  In the long run,
on the time scale much longer that $1/\Gamma_R$, the population of mode $b_2$ drops
to the single-atom level (this is seen also in Fig. \ref{FG2}, where it happens on
the short time scale due to a small $\Gamma$).

In the linear case Eq. (\ref{EQ45}) acts like a dispersive beam-splitter (see, for
example, Ref. \cite{mogilevtsev2009}). Thus, a strong loss in the central well
induces quantum correlations between the side wells, i.e. for times $t\gg
1/\Gamma_R$ the cold atoms occupy the state
\eqb
|\Psi_n\rangle \sim (b_2^\dagger)^n|0\rangle\sim
(a_1^{\dagger}-a_2^{\dagger})^n|0\rangle
\eqe
which is unaffected by the dissipation described by Eq. (\ref{EQ45}).

In the nonlinear case, the  dissipation of the $\pi$-phase mode $b_2$  is
surprisingly non-trivial. This can be clearly demonstrated in the case when the
decay of the zero-phase mode $b_1$ occurs on the faster scale than the inter-mode
dynamics. To uncover the details, let us use the higher-order validity of Eq.
(\ref{EQ45}) and reduce it to a single-mode master equation for $b_2$, by assuming
that $U\langle n_2\rangle/\hbar\ll \Gamma_R$ and $J_{12}/\hbar \ll \Gamma_R$.


\section{The reduced master equation for $\pi$-phase coherent  mode }
\label{sec3}
We now reduce the master equation (\ref{EQ45}) to that for the mode $b_2$ alone, in
the case of a strong dissipation of mode $b_1$ as compared to the Hamiltonian
dynamics of the coherent  modes $b_1$ and $b_2$. The derivation is similar to that
of the previous section. First, we pass to the interaction representation:
\begin{eqnarray}
&& \hat{H}_{II}(t) = U^\dag(t) H_{II} U(t),\quad \hat{\rho}_{12}= U^\dag (t)\rho_{12}(t)
U(t)\nonumber\\
&& U(t) = U_{1}(t)\otimes U_{2}(t)=\exp\{-\frac{i}{\hbar}(H_{1}\! +\!H_2)\Delta
t\},
\label{EQ46}\end{eqnarray}
where the respective Hamiltonian terms, derived from Hamiltonian (\ref{EQ3}) with
account of the Lamb shift (\ref{EQ35}), read:
\[
H_1 = \vare_R n_1 +\frac{U}{2}n^2_1, \quad H_2 =\frac{U}{2}n^2_2,
\]
\eqb
H_{II} = - J_{12}(b^\dag_1b_2+b^\dag_2b_1) +2Un_1n_2 +
\frac{U}{2}[b^\dag_1b_2]^2+\frac{U}{2}[b^\dag_2b_1]^2.
\label{EQ48}\eqe

Introducing the small parameter $\epsilon^\prime =
\mathcal{O}(J_{12}/\hbar\Gamma_R)$, and  assuming that $U\langle
n_{2}\rangle/(\hbar\Gamma_R))=\mathcal{O}(\epsilon^\prime)$, one can derive the
equations for the density matrix in the interaction representation, which turn out
to be similar to those of the previous section (see Eqs. (\ref{EQ21})-(\ref{EQ24}))
with the obvious changes. The form of the density matrix $\rho^{(1)}_{12}$ is also
similar to that of Eqs. (\ref{EQ22}) and (\ref{EQ26}):
\eqb
\rho^{(1)}_{12}(t) = W(t)\left[\rho^{(0+1)}_{2}\otimes|0\rangle\langle0|
\right]W^\dag(t) +  \mathcal{O}(e^{-\frac{\Gamma_1}{2}\Delta t}),
\label{EQ49}\eqe
where $W(t)$ reads
\eqb
W(t) = I + s_1(t)b_2(t)b^\dag_1(t)+ s_2(t)b_2^2(t)[b^\dag_1(t)]^2,
\label{EQ50}\eqe
with some scalar functions $s_j(t)=\mathcal{O}(\epsilon^\prime)$. Using the same
routine as in the previous section, one obtains the equations for the parameters
$s_j(t)$:
\begin{subequations}
\label{EQ51}
\begin{eqnarray}
&&\frac{ds_1}{dt} = -\left[\frac{\Gamma_R}{2}+i\frac{\vare_R}{\hbar}\right]s_1 +
i\frac{J_{12}}{\hbar},\\
&&\frac{ds_2}{dt} = -\left[\Gamma_R+i\frac{2\vare_R}{\hbar}\right]s_2 - i\frac{U}{2\hbar}.
\end{eqnarray}
\end{subequations}
These can be easily solved to give:
\begin{eqnarray*}
&& s_1 = \frac{2iJ_{12}}{\hbar\Gamma_R}\left[1+i\frac{2\vare_R}{\hbar\Gamma_R}\right]^{-1}
+\mathcal{O}(e^{-\frac{\Gamma_R}{2}\Delta t}),\\
&& s_2 =
-\frac{iU}{2\hbar\Gamma_R}\left[1+i\frac{2\vare_R }{\hbar\Gamma_R}\right]^{-1}
+\mathcal{O}(e^{-\frac{\Gamma_R}{2}\Delta t}).
\end{eqnarray*}
Define the following parameters:
\eqb
\Gamma_j =
\frac{K_j^2}{\hbar^2\Gamma_R}\left[1+\left(\frac{2\vare_R}{\hbar\Gamma_R}\right)^2\right]^{-1},
\; \varkappa_j=-\vare_R\frac{\Gamma_j}{\Gamma_R},
\label{G12}
\eqe
where $K_1 = 2J_{12}$ and $K_2 = U$.  Then, in a similar way as in the previous
section, by taking the partial trace over the subspace of mode $b_1$, one derives a
closed  master equation for mode $b_2$. In the Schr\"odinger picture we have
\eqb
 \frac{d\rho_{2}}{dt} =
-\frac{i}{\hbar}[\tilde{H}_2,\rho_{2}] +
\Gamma_1D_1\{\rho_2\}+\Gamma_2D_2\{\rho_2\}
+\mathcal{O}(e^{-\frac{\Gamma_R}{2}\Delta t}),
\label{EQ52}\eqe 
where the Hamiltonian,  augmented by the Lamb shifts, and the two dissipation
channels read:
\eqb
\tilde{H}_2 = \frac{U}{2}n_2^2 +\varkappa_1n_2+\varkappa_2n_2(n_2-1),
\label{EQ53}\eqe
\eqb
D_1\{\rho\} = b_2\rho b_2^\dag - \frac12n_2\rho -\frac12 \rho n_2,
\label{EQ54}\eqe
\eqb
D_2\{\rho\} = b_2^2\rho (b_2^\dag)^2 - \frac12(b_2^\dag)^2b_2^2\rho -\frac12 \rho
(b_2^\dag)^2b_2^2.
\label{EQ55}\eqe
Finally, let us gather together the conditions used in derivation of the reduced
equation (\ref{EQ52}). We have
\eqb
\frac{UN}{\hbar},\;\frac{J_{12}}{\hbar}\ll\Gamma_R\sim\frac{J^2_{13}}{\hbar^2
\Gamma},\quad \frac{J_{13}}{\hbar \Gamma} \ll  1.
\label{EQ59}\eqe

The validity conditions (\ref{EQ59}) can be recast in terms of the characteristic
tunneling times and the nonlinear time. Defining $\tau_{13} = \hbar/J$, $\tau_{12}
= \hbar/J_{12}$ and $\tau_{nl} = \hbar/U$, we have: $\frac{\tau_{13}}{\tau_{12}},\;
\frac{N\tau_{13}}{\tau_{nl}}\ll (\Gamma\tau_{13})^{-1}\ll 1$. The rates of the  two
dissipation channels of mode $b_2$  have the following orders:  $\Gamma_1\sim
(\tau_{13}/\tau_{12})^2/\Gamma$ and $\Gamma_2 \sim (\tau_{13}/\tau_{nl})^2/\Gamma$.

We note that a master equation similar to Eq. (\ref{EQ52}) has already appeared
before in connection with one- and two-photon absorption in quantum optics, where
its special cases were studied \cite{BosonPair}. It was shown that two-particle
absorption has properties drastically different from the single-particle one. In
particular, the decay is non-exponential and, irrespectively of the number of
particles in the initial state of the mode $b_2$, number of particles in this mode
drops to the single-particle level during the same time-interval \cite{BosonPair}.

\begin{figure}[htb]
\begin{center}
\epsfig{file=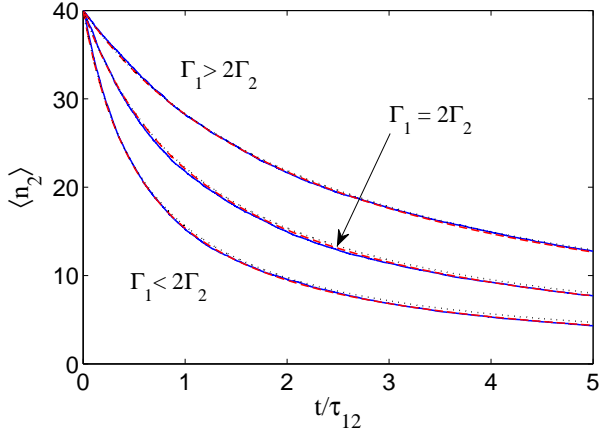,width=0.45\textwidth} \caption{(Color online) Comparison of
the two-mode master equation (\ref{EQ45}) with the single-mode master equation
(\ref{EQ52}) and the analytical approximation of Eq.  (\ref{EQn3}). The solid lines
give the two-mode equation, the dashed gives the single-mode, and the dotted -- the
analytical approximation. Here: $N=40$, $\Gamma_R\tau_{12} = 200$, and, from the
top to bottom, $\Lambda \equiv {UN}/{2J_{12}} = 20$, $N/\sqrt{2}$, $40$. To compare
the results in the domain of their validity, i.e. after the atoms are removed from
mode $b_1$, we have used the Fock initial state with all atoms occupying mode $b_2$
(a Gaussian with a small dispersion leads to the same results).}
\label{FG3}
\end{center}
\end{figure}

Eq. (\ref{EQ52})  has a number of specific  features. First, we see that in the
symmetric potential (when $\Gamma_1=0$) the decay occurs due to loss of two
particles at once and the quantum parity, being average of the quantum parity
operator
\eqb
P=(-1)^{b_2^\dag b_2}
\eqe
remains constant. For example, for the state with the $\langle P(0) \rangle=-1$
(odd parity), one will have $\langle P(t) \rangle=-1$; the superposition state with
only the odd (even) number of atoms will remain the state with the odd (even)
number of atoms during all the evolution time. Second, for a biased potential
($\Gamma_1\ne 0$) there is a resonance between the two different dissipation
channels, under the condition $\Gamma_1=2\Gamma_2$, resulting in a polynomial decay
of population. To see this, consider evolution of the average population
\eqb
\frac{d \langle n_2\rangle}{dt} = -(\Gamma_1-2\Gamma_2)\langle n_2\rangle -
2\Gamma_2\langle n_2\rangle^2 -2\Gamma_2 (\Delta n_2)^2,
\label{EQn2}\eqe
where $(\Delta n_2)^2 = \langle n_2^2\rangle - \langle n_2\rangle^2$. The initial
state of mode $b_2$, to be used in Eq. (\ref{EQ52}), is a Fock state with a good
approximation (mode $b_1$ is emptied on a much faster time scale). Hence,
discarding $\Delta n_2$ (which is justified by numerical simulations, see Fig.
\ref{FG3}), we get an approximation
\eqb
 \langle n_2(t)\rangle \approx \frac{\langle n_2(t_0)\rangle e^{-\gamma (t-t_0)}}{1+
\frac{2\Gamma_2}{\Gamma_1-2\Gamma_2} \langle
n_2(t_0)\rangle[1-e^{-(\Gamma_1-2\Gamma_2)(t-t_0) }]}
\label{EQn3}\eqe
giving,  for $\Gamma_1=2\Gamma_2$, the $t^{-1}$-decay: $\langle n_2(t)\rangle^{-1}
\approx \langle n_2(t_0)\rangle^{-1} + \Gamma_1(t-t_0) $. Thus, we have  a quantum
resonance between two different (linear and non-linear) dissipation channels of a
subsystem (mode $b_2$). The matching between them is expressed in terms of matching
between the linear bias and non-linear interaction coefficient: $U =
\sqrt{2}J_{12}=(\vare_2-\vare_1)/\sqrt{2}$.  In Fig. \ref{FG3} we show an excellent
agreement of the analytical approximation, Eq. (\ref{EQn3}), with the numerical
simulations, i.e. the Monte Carlo wave-function method \cite{MCD}, of the
single-mode (\ref{EQ52}) and the two-mode (\ref{EQ45}) master equations.

\section{The mean-field approximation}
\label{sec4}
The quantum derivations of the reduced master equations  from the first principles,
presented above,  are  involved.  One then may inquire if the mean-field
approximation, commonly applied to the many-boson models with large number of
bosons and which is much simpler to analyze,  can substitute the quantum
derivation. Here we remind that the two-site model with a local dissipation is
perfectly described by the mean-field approximation \cite{TWW,NlZeno}. This,
however, turns out to be not the case   for  the three-site Bose-Hubbard model, as
we will show below.

The mean-field Hamiltonian can be obtained from Hamiltonian (\ref{EQ3}) by
replacing the boson operators with $c$-numbers ($b_j\to\beta_j$) \cite{BStat}, we
get
\[
\mathcal{H} = -J_{13}(\beta^*_1\beta_3 +\beta_3^*\beta_1)
-J_{12}(\beta^*_1\beta_2+\beta^*_2\beta_1) +\vare|\beta_3|^2
\]
\eqb
+U|\beta_3|^4 +\frac{U}{2}\left([|\beta_1|^2 +|\beta_2|^2]^2 +
[\beta^*_1\beta_2+\beta^*_2\beta_1]^2\right).
\label{MF1}\eqe
Note that the total number of atoms is given as $\sum_{j=1}^3|\beta_j|^2 = N$. The
local atom loss (dissipation) part of Eq. (\ref{EQ1}) can be simply added to the
mean-field Hamiltonian equations as an atom loss of mode $\beta_3$  (as we will see
shortly, it is describable classically; see also Refs. \cite{TWW,NlZeno}), that is
\eqb
\frac{d\beta_j}{dt} = -\frac{i}{\hbar}\frac{\pd \mathcal{H}}{\pd \beta^*_j}
-\delta_{j,3}\frac{\Gamma}{2}\beta_3.
\label{MF2}\eqe
Thus, the mean-field equations read:
\begin{subequations}
\begin{eqnarray}
 \frac{d\beta_1}{dt} &=& \frac{iJ_{13}}{\hbar}\beta_3 + \frac{iJ_{12}}{\hbar}\beta_2 -\frac{iU}{\hbar}(|\beta_1|^2
+2|\beta_2|^2)\beta_1 \nonumber\\
&& -\frac{iU}{\hbar}\beta^2_2\beta^*_1,
\label{EQb1}\\
\frac{d\beta_2}{dt} &=& \frac{iJ_{12}}{\hbar}\beta_1 -\frac{iU}{\hbar}(|\beta_2|^2
+2|\beta_1|^2)\beta_2 -\frac{iU}{\hbar}\beta^2_1\beta^*_2,\qquad
\label{EQb2}\\
 \frac{d\beta_3}{dt} &=& \frac{iJ_{13}}{\hbar}\beta_1
-\left(\frac{\Gamma}{2}+\frac{i\vare}{\hbar}\right)\beta_3
-\frac{2iU}{\hbar}|\beta_3|^2\beta_3.
\label{EQb3}
\end{eqnarray}
\end{subequations}

\subsection{The first reduction: equations for $\beta_1$ and $\beta_2$}

Consider the strongly dissipated case of section \ref{sec2}. The small parameter is
$\epsilon = \frac{J_{13}}{\hbar\Gamma}$ with the conditions
$\frac{J_{12}}{\hbar\Gamma} = \mathcal{O}(\epsilon)$ and $\frac{UN}{\hbar\Gamma} =
\mathcal{O}(\epsilon)$. For $t\gg 1/\Gamma$, $\beta_3 = \mathcal{O}(\epsilon)$ and
one can integrate Eq. (\ref{EQb3}) by rewriting it in the integral form and
neglecting the higher-order nonlinear term. Using the integration by parts, we get
\eqb
\beta_3(t) = \frac{2iJ_{13}}{\hbar\Gamma}\left(1+
\frac{2i\vare}{\hbar\Gamma}\right)^{-1}\beta_1(t)
+\mathcal{O}(\epsilon^2)+\mathcal{O}(e^{-\frac{\Gamma \Delta t}{2}}).
\label{EQ56}\eqe
This result corresponds to the expression for the full density matrix (\ref{EQ25})
with Eqs. (\ref{EQ26}) and (\ref{EQ33}), where the  amplitude  $\beta_3$ is locked
to that of  $\beta_1$ and $\beta_2$ with the same coefficient as in the full
quantum case  ($\alpha_1$ of Eq. (\ref{EQ33})). Eq. (\ref{EQ56}) can be now
inserted into Eqs. (\ref{EQb1}) and (\ref{EQb2}). We obtain a reduced system
describing the coherent modes:
\begin{subequations}
\begin{eqnarray}
\frac{d\beta_1}{dt} &=&
-\left(\frac{\Gamma_R}{2}+\frac{i\vare_R}{\hbar}\right)\beta_1
+\frac{iJ_{12}}{\hbar}\beta_2  - \frac{iU}{\hbar}\beta^2_2\beta^*_1 \nonumber\\
& -&\frac{iU}{\hbar}(|\beta_1|^2
+2|\beta_2|^2)\beta_1+\mathcal{O}(\epsilon^3)+\mathcal{O}(e^{-\frac{\Gamma\Delta
t}{2}}),
\label{EQb12A}\\
\frac{d\beta_2}{dt} &=& \frac{iJ_{12}}{\hbar}\beta_1 - \frac{iU}{\hbar}(|\beta_2|^2
+2|\beta_1|^2)\beta_2 - \frac{iU}{\hbar}\beta^2_1\beta^*_2,\qquad
\label{EQb12B}
\end{eqnarray}
\end{subequations}
where $\Gamma_R$ is given by   Eq. (\ref{EQ18}) and the Lamb shift $\vare_R$ by Eq.
(\ref{EQ35}) (i.e., by the corresponding quantum results).

\subsection{The second reduction: equation for $\beta_2$}

Now, let us perform the second reduction to an equation for the amplitude
$\beta_2$, similar  as in the quantum case of section \ref{sec3}. In section
\ref{sec3} we have assumed that the new small parameter is $\epsilon^\prime =
\frac{J_{12}}{\hbar\Gamma_R}$ with the additional condition on the nonlinearity
$\frac{U\langle n_2\rangle}{\hbar\Gamma_R} = \mathcal{O}(\epsilon^\prime)$.
However, let us for a while broaden the derivation and discard the   condition on
the nonlinearity. For times $t\gg 1/\Gamma_R$  boson mode $\beta_1$ is practically
empty, i.e. $\beta_1 =\mathcal{O}(\epsilon^\prime)$. This allows us to simplify
Eqs. (\ref{EQb12A}) and (\ref{EQb12B}) as follows
\begin{subequations}
\begin{eqnarray}
\frac{d\beta_1}{dt} &=&
-\left[\frac{\Gamma_R}{2}+\frac{i\vare_R}{\hbar}\right]\beta_1
+\frac{iJ_{12}}{\hbar}\beta_2  - \frac{iU}{\hbar}[\beta^2_2\beta^*_1+2|\beta_2|^2\beta_1 ]\nonumber\\
&&+\mathcal{O}({\epsilon^\prime}^3)+\mathcal{O}(e^{-\frac{\Gamma_R\Delta t}{2}}),
\label{Eb1R}\\
\frac{d\beta_2}{dt} &=& \frac{iJ_{12}}{\hbar}\beta_1 -
\frac{iU}{\hbar}|\beta_2|^2\beta_2 -
\frac{iU}{\hbar}\beta^2_1\beta^*_2+\mathcal{O}({\epsilon^\prime}^3)\nonumber\\
&&+\mathcal{O}(e^{-\frac{\Gamma_R\Delta t}{2}}),\qquad
\label{Eb2R}
\end{eqnarray}
\end{subequations}
where, for simplicity, we have dropped the term $\mathcal{O}(\epsilon^3)$.  Already
from this system it is clear that the mean-field approach will not account for the
boson-pair dissipation channel (\ref{EQ55}) of mode $b_2$, present  in the full
quantum Eq. (\ref{EQ52}). Indeed, Eq. (\ref{Eb1R}) must be somehow integrated with
the result to be inserted into Eq. (\ref{Eb2R}). However, one can notice that the
parameter $U$ enters Eq. (\ref{Eb1R}) only in the conjunction with a factor
$\sim|\beta_2|^2$, hence all the terms in Eqs. (\ref{Eb1R}) and (\ref{Eb2R}) scale
as $\sqrt{N}$ (the scale of $\beta$) if we fix the nonlinearity parameter
$\Lambda=UN$, whereas in the quantum case the $N$-scaling of the boson-pair channel
is $\mathcal{O}(1)$.

Now, under  the   condition on the nonlinearity as in the full quantum case, i.e.
$\frac{U\langle n_2\rangle}{\hbar\Gamma_R} = \mathcal{O}(\epsilon^\prime)$,  one
can proceed to derive the reduced equation for the amplitude $\beta_2$. To this
end, the system of equations for $\beta_1$ and $\beta^*_1$ in the required order
can be written in the following form
\eqb
\frac{d}{dt}\left(\begin{matrix} \beta_1\\ \beta^*_1\end{matrix}\right) =
\frac{\Gamma_R}{2}M\left(\begin{matrix}\beta_1\\
\beta^*_1\end{matrix}\right)+\frac{iJ_{12}}{\hbar}\left(\begin{matrix}\beta_2\\-\beta^*_2\end{matrix}\right),
\eqe where
\eqb
M =\left(\begin{matrix}-\gamma-\frac{4iU}{\hbar\Gamma_R}|\beta_2|^2 &
-\frac{2iU}{\hbar\Gamma_R}(\beta_2)^2\\ \frac{2iU}{\hbar\Gamma_R}(\beta^*_2)^2 &
-\gamma^*+\frac{4iU}{\hbar\Gamma_R}|\beta_2|^2
\end{matrix}\right),
\label{EQ60}\eqe
with $\gamma = \left(1+\frac{2i\vare_R}{\hbar\Gamma_R}\right)$. Eq. (\ref{EQ60})
can be put into the integral form and integrated for times $t\gg1/\Gamma_R$, in
this case the matrix $\frac{\Gamma_R}{2}M(t)$ enters the exponent under the
integral, with the result
\eqb
\left(\begin{matrix} \beta_1\\ \beta^*_1\end{matrix}\right) =
-\frac{2iJ_{12}}{\hbar\Gamma_R}M^{-1}\left(\begin{matrix} \beta_2\\
-\beta^*_2\end{matrix}\right) +\mathcal{O}(e^{-\frac{\Gamma_R\Delta
t}{2}})+\mathcal{O}({\epsilon^\prime}^2).
\label{EQ61}\eqe
We need only the first row of the matrix $M^{-1}$:
\eqb
(M^{-1})_{11} = -\gamma^{-1} + \gamma^{-2}\frac{4iU}{\hbar\Gamma_R}|\beta_2|^2
+\mathcal{O}({\epsilon^\prime}^2),
\label{EQ62}\eqe
\eqb
(M^{-1})_{12} = -|\gamma|^{-2}\frac{2iU}{\hbar\Gamma_R}(\beta_2)^2
+\mathcal{O}({\epsilon^\prime}^2).
\label{EQ63}\eqe
From  Eqs. (\ref{EQ61})-(\ref{EQ63}) we obtain
\eqb
\beta_1 = \gamma^{-1}\frac{2iJ_{12}}{\hbar\Gamma_R}\beta_2
+\mathcal{O}({\epsilon^\prime}^2) +\mathcal{O}(e^{-\frac{\Gamma_R\Delta t}{2}}).
\label{EQ64}\eqe
Inserting this expression into Eq. (\ref{Eb2R}) we arrive at the reduced aquation
for the amplitude $\beta_2$, the mean-field analog  of the coherent $\pi$-phase
mode $b_2$:
\begin{eqnarray}
\frac{d\beta_2}{dt} &=&
-\left(\frac{\Gamma_1}{2}+\frac{i\varkappa_1}{\hbar}\right)\beta_2
-\frac{iU}{\hbar}|\beta_2|^2\beta_2
\nonumber\\
&&+\mathcal{O}({\epsilon^\prime}^3) +\mathcal{O}(e^{-\frac{\Gamma_R\Delta t}{2}}),
\label{EQ65}\end{eqnarray}
with $\Gamma_1$ and $\varkappa_1$ given in Eq. (\ref{G12}). Observe that, while the
single-boson dissipation channel, Eq. (\ref{EQ54}), is accounted by the mean-field
Eq. (\ref{EQ65}) (the first term on the right hand side), the boson-pair channel,
Eq. (\ref{EQ55}), is not.

In conclusion of this section, we have shown  that, while the single-boson
dissipation channel of the  coherent  mode $b_2$ can be described  by the
mean-field approach, the boson-pair dissipation channel cannot be captured by the
mean-field approximation and, thus, it has   \textit{quantum nature}.


\section{Discussion of the results}
\label{sec5}

We have considered the derivation of the reduced master equations in the limit of
strong dissipation on the example of the  Bose-Hubbard model with a local external
dissipation (i.e., the atom loss from the central  site). The method we have used
is not based on the assumption of the factorization of the full density matrix,
instead we demonstrate that one can effectively solve  the master equation directly
in the lowest orders of a small parameter (inversely proportional to the local
dissipation rate). On this way, one is able to obtain the reduced master equations
for the subsystems (the coherent modes) of the higher-order in the small parameter
(e.g., we have derived the equation up to the third order).

The derivation reveals the following features. First of all, the full density
matrix is not factorized (which is the usual assumption, see for instance Ref.
\cite{BP}) but is expressed in the form of a ``dressed'' factorized density matrix,
where the population of the strongly dissipated mode depends on that of the other
modes. Nevertheless, the reduced density matrix of a subsystem is shown to satisfy
a master equation in the Lindblad form.  This feature appears in the two reduced
master equations derived in the present paper, thus suggesting an universality.
Moreover, the Lamb shifts and the dissipation rates of the subsystem turn out to be
given by the similar expressions in the two cases, suggesting even the universality
of the form of expressions for these quantities.

We have analyzed  the relation between the full quantum derivation of the reduced
master equation for the density matrix of a subsystem and the mean-field adiabatic
elimination procedure. We have found that the mean-field approximation applied to
the Bose-Hubbard model cannot capture all dissipation channels  of a subsystem,
even if  the   external dissipation applied to the   system is describable
classically (in our case, the  single-boson dissipation channel describing the
removal of atoms). Namely, in the three-site Bose-Hubbard model the $\pi$-phase
coherent mode has the boson-pair dissipation channel, which is not captured by the
mean-field approximation, and the single-boson dissipation, captured by it. This is
a quite distinct situation from the two-site model, where the dissipation dynamics
is described by the mean-field approximation with a good accuracy
\cite{TWW,NlZeno}. Here we note, however,  that in the two-site  model with
dissipation the boson-pair tunneling is not possible. Though we have considered the
three-site Bose-Hubbard model, the boson-pair dissipation channel is a
\textit{general feature} of the multi-site models, since it is the result of a
nonlinearity of the model and the fact that more than one filled site is coupled to
the  dissipated site, which serves as a common reservoir.

The failure of the mean-field approximation to the quantum master equation means
that the dissipation channels not accounted by the mean-field have a
\textit{genuine quantum nature}. This imposes severe limitations on the
applicability of the semiclassical mean-field approach to the Bose-Hubbard models
considered as open quantum systems, even when the external dissipation is
describable classically.  Invoking the link to the discrete WKB approach, mentioned
in the Introduction, one has to  develop a more general approach by going to the
first-order approximation  in the effective Planck constant $1/N$, i.e. a
$1/N$-correction to the mean-filed equations,  to capture the boson-pair
dissipation channel. This is an important conclusion, in view of the current
experiments on the single-site addressability and controlled measurement via the
local atom losses of Bose-Einstein condensate loaded in the optical lattices
\cite{SiteAddr1,SiteAddr2}, describable by the open multi-site Bose-Hubbard models.

\end{document}